\documentclass[conference,a4paper]{IEEEtran}

\usepackage[table, dvipsnames]{xcolor}
\usepackage{xcolor,colortbl}
\newcolumntype{a}{>{\columncolor{gray!30}}c}

\usepackage{soul}
\usepackage{enumitem}
\usepackage{comment}
\usepackage{array}
\usepackage{cite}
\usepackage{amsmath,amssymb,amsfonts}
\usepackage{algorithmic}
\usepackage{graphicx}
\usepackage{textcomp}
\usepackage{soul}
\usepackage{graphicx, caption, subcaption}
\usepackage{multirow}
\usepackage{multicol}
\usepackage{array}
\usepackage{physics}
\usepackage{subfig}
\usepackage[normalem]{ulem}

\newcount\Comments  
\definecolor{darkgreen}{rgb}{0,0.5,0}
\definecolor{darkred}{rgb}{0.7,0,0}
\definecolor{teal}{rgb}{0.1,0.6,0.7}
\definecolor{blue}{rgb}{0.0,0.1,0.9}
\definecolor{orange}{rgb}{1.,0.7,0.0}
\definecolor{lightblue}{rgb}{0.70, 0.80, 0.89}

\newcommand{\kibitz}[2]{\ifnum\Comments=1{{\textcolor{#1}{\textsf{#2}}}}\fi}

\newcolumntype{C}[1]{>{\centering\arraybackslash}p{#1}}
\def\BibTeX{{\rm B\kern-.05em{\sc i\kern-.025em b}\kern-.08em
    T\kern-.1667em\lower.7ex\hbox{E}\kern-.125emX}}

\newcommand\blfootnote[1]{%
\begingroup
\renewcommand\thefootnote{}\footnote{#1}%
\addtocounter{footnote}{-1}%
\endgroup
}

\makeatother

\IEEEoverridecommandlockouts

\begin{document}

\title{
Quantum Computing for Large-scale Network Optimization: Opportunities and Challenges}

\author{
\IEEEauthorblockN{Sebastian Macaluso, Giovanni Geraci,
Elías F. Combarro,\\
Sergi Abadal,
Ioannis Arapakis,
Sofia Vallecorsa and
Eduard Alarcón}
}

\bstctlcite{IEEEexample:BSTcontrol}

\maketitle
\blfootnote{This work has been accepted for publication in \textit{IEEE Communications Magazine}.}
\blfootnote{The work of Giovanni Geraci was in part supported by the Spanish Research Agency through grants PID2021-123999OB-I00 and CNS2023-145384 and by the Maria de Maeztu Units of Excellence Programme (CEX2021-001195-M).
The work of Elías F. Combarro was partially supported by grant PID2023-146520OB-C22, funded by MICIU/AEI/10.13039/501100011033, by grant IDE/2024/000734 funded by Principado de Asturias, and by the Ministry for Digital Transformation and of Civil Service of the Spanish Government through the QUANTUM ENIA project call – Quantum Spain project, and by the European Union through the Recovery, Transformation and Resilience Plan – NextGenerationEU within the framework of the Digital Spain 2026 Agenda.
The work of Eduard Alarcón was partially supported by the ICREA Academia Award 2024.}
\begin{abstract}

The complexity of large-scale 6G-and-beyond networks demands innovative approaches for multi-objective optimization over vast search spaces, a task often intractable. Quantum computing (QC) emerges as a promising technology for efficient large-scale optimization. We present our vision of leveraging QC to tackle key classes of problems in future mobile networks. By analyzing and identifying common features, particularly their graph-centric representation, we propose a unified strategy involving QC algorithms. Specifically, we outline a methodology for optimization using quantum annealing as well as quantum reinforcement learning. Additionally, we discuss the main challenges that QC algorithms and hardware must overcome to effectively optimize future networks.

\end{abstract}

\section{Introduction} 
\label{sec:Intro}

Quantum computing (QC) has rapidly emerged as a promising field, with its unparalleled potential to tackle problems typically intractable for classical computers. Quantum bits (qubits) leverage the principles of superposition, interference and entanglement to accelerate computations and open the door to previously unimaginable algorithms. This fundamental characteristic allows quantum computers to perform complex calculations at speeds exponentially faster than their classical counterparts in certain domains, enabling breakthroughs in fields such as cryptography, materials science, and artificial intelligence (AI).

Developments in QC pave the way for novel solutions to intractable optimization problems and are
expected to play a disruptive role in multiple industries.
For example, optimizing supply chain logistics, financial portfolios, and improving manufacturing processes could benefit from quantum speedups. However, the application of QC to mobile communications remains a largely unexplored area. 
As 6G-and-beyond networks are expected to grow in size and complexity with heterogeneous nodes, additional frequency bands, and multi-technology coexistence requirements, could quantum computing drive their large-scale optimization?


There are three main areas of telecommunications where quantum technologies can have a tremendous impact: quantum communications, 
quantum cryptography, and QC. We focus on QC, introducing opportunities and challenges associated with its application to the large-scale optimization of mobile networks. 
While recent work has mainly given an overview of the potential benefits of QC for wireless networks \cite{QML46G,9870532, 10233127,
kim2021heuristic,duong2022quantum}, in this paper we provide detailed methodologies for leveraging QC,
delving into the specifics of how QC can be practically applied and architecturally integrated to tackle complex network optimization problems.

It is worth noting that, to the best of our knowledge, there are no general quantitative statements on the performance of quantum approaches for combinatorial optimization problems. Current understanding is limited both theoretically and practically. While we cannot yet assert a clear advantage in using QC for large problems, ongoing advancements in hardware and algorithms increase the potential for improvements in efficiency, solution quality, and energy use. Further research is needed to clarify this, which we aim to motivate through concrete examples and directions in our paper.

\section{A Primer on Quantum Computing} 
\label{sec:QC}

Quantum computing is a new computational paradigm that explicitly exploits properties of quantum systems 
to obtain advantages over classical computers. Paramount examples 
include Shor's algorithm for factoring large integers
and Grover's procedure for finding specific elements in unsorted databases.

In the last few years, researchers have focused on potential advantages from present-day quantum computers, usually called Noisy Intermediate Scale Quantum (NISQ) devices.
%
%
These computers have a reduced number of qubits, 
are subject to noise and errors in both operations and measurements, and present limited connectivity among their qubits. Thus, NISQ devices necessitate
error correction to convert physical qubits into logical ones. 
Moreover, due to a phenomenon called decoherence, qubits lose their quantum properties in a short time, reducing the number of operations that can be performed on them, before decoherence corrupts the information.
Despite NISQ devices limitations, they have shown quantum advantage in artificial, academic tasks. Also, there is theoretical evidence that quantum computers may outperform classical methods for certain kinds of problems~\cite{pirnay} or, at least, constitute viable alternatives to other heuristic algorithms.

\subsection*{\bf Quantum Computer Categories}

Quantum computers
can be divided in two different categories, as follows.  

\subsubsection*{Analog QC}

These work by evolving the state of their qubits in a continuous way.
The most popular type 
are quantum annealers (QAs), based on quantum adiabatic computing. QAs
reach the highest qubit count among all currently available quantum computers---5000+ in D-Wave's Advantage model, with 7000+ expected for the recently announced Advantage2---%
but
the applicability and flexibility of these machines are reduced. 

\subsubsection*{Digital QC}

Digital quantum computers apply discrete, unitary transformations (called quantum gates) to the state of the qubits.
They are capable of universal quantum computation and, thus, are not restricted to solving optimization problems as in the case of QAs. 

\begin{figure}[!t]
\centering
\includegraphics[width=\columnwidth]{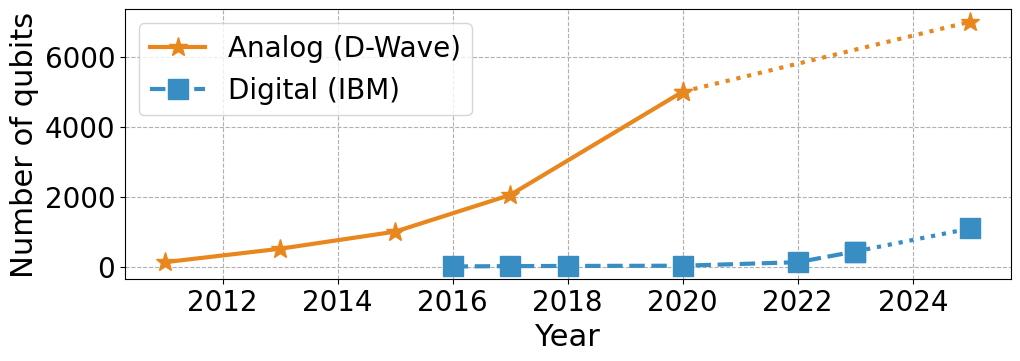}
\caption{Timeline for the evolution of number of qubits of leading developers.
The dotted line indicates expected releases.
The two curves are shown in the same figure but cannot be compared to each other, as the applicability and the way to implement computations is fundamentally different for each (digital or analog) technology.}
\label{fig:vision}
\vspace{-0.2cm}
\end{figure}

We show in Fig.~\ref{fig:vision} the evolution of the number of qubits on leading analog (D-Wave) and digital (IBM) QC developers. 
The number of qubits to perform calculations is the main metric, but
error mitigation and coherence time of a qubit are important aspects as well. Also, there is no winning hardware technology at the moment. The main ones include superconducting qubits, trapped-ion qubits, photonics, and neutral atoms.

\subsection*{\bf Quantum Optimization}

The main use of quantum computers in optimization comes from the possibility of transforming almost any combinatorial optimization problem
into an instance of finding the ground state of a quantum Hamiltonian, i.e., the state of minimum energy.
The Hamiltonian is an operator that acts on the qubits, depends on their state and interactions, and represents the total energy of this system.
The Hamiltonian of choice is usually the Ising one, with quadratic terms from the interaction between adjacent sites (qubits) in a graph.
However, a more natural way of reformulating these problems is the Quadratic Unconstrained Binary Optimization (QUBO) formalism, in which the problems take the form of minimizing a quadratic polynomial on binary variables. 

Once the problem is formulated as either a QUBO or an Ising model instance, it can be solved with analog quantum computers via quantum annealing (QA) or with digital ones via the Quantum Approximate Optimization Algorithm (QAOA) (or the Variational Quantum Eigensolver).
Also, we note that QAOA can be applied more in general with the Polynomial Unconstrained Binary Optimization (PUBO) formalism, which does not require the quadratic constraint from QUBO.
As we present later, these techniques could be applied to perform cellular RAN deployment optimization.

\subsection*{\bf  Quantum Machine Learning}

Quantum machine learning (QML)
emerges at the intersection of QC and classical machine learning, aiming to improve the efficiency and effectiveness of learning models by using quantum information processing. 
One of the most popular QML methods involves the creation of purely quantum models based on quantum architectures with free trainable parameters (e.g., variational circuits or quantum Boltzmann machines), giving rise to 
quantum neural networks.

There is a wide variety of tasks where quantum neural networks can be applied, ranging from supervised to unsupervised learning and even reinforcement learning (RL) problems. 
Quantum computers can enhance the performance of deep RL, particularly where the state-action space is large.
For instance, quantum generalizations of classical energy-based models inspired by statistical physics (e.g. restricted Boltzmann machines), provide speed-ups for deep energy-based RL in large spaces~\cite{PRXQuantum.2.010328}, and some of these algorithms can be run on current NISQ devices.
Also, free energy-based reinforcement learning (FERL) with clamped quantum Boltzmann machines (QBM) significantly improves the learning efficiency compared to classical Q-learning on discrete  and continuous  state-action spaces \cite{Schenk_2024}. 
Thus, evidence shows that QML could improve performance and learn faster or with fewer examples.
As we detail below, QML can offer novel ways to tackle mobility management and virtual network function scheduling in future networks.

\section{Graph-centric Optimization Problems in Next-generation Mobile Networks} 
\label{sec:vision}

Being qualitatively different in nature to classical computing, QC has the potential to  disrupt the way we tackle large-scale optimization. 
We expect QC to provide an advantage with better quality or more efficient solutions compared to traditional methods on complex scenarios involving high dimensional heterogeneous spaces of configurations typical of large-scale networks. We do not foresee an improvement of the same extent over classical algorithms on small-scale networks optimization problems.
In this section, we describe three representative classes of  problems in mobile networks, highlighting their large-scale, graph-centric nature. 
We stress this graph-centric nature, given the potential of QC to perform efficient exploration over the graph's structure.
Thus, for each class of problem, we discuss current approaches and their limitations, and schematically show their mapping to a graph structure.



\subsection*{\bf Cellular RAN Deployment Optimization}

The coverage and capacity of cellular networks are significantly influenced by the deployment sites of cells and the configuration of base station antennas. Optimizing these parameters, a process known as cell shaping, is inherently challenging. The settings across cells are coupled by interference, making the optimization problem non-convex and NP-hard. 
Radio resource allocation, already a large-scale optimization problem, is set to become even more complex with the introduction of large antenna arrays, due to the increased degrees of freedom spanning time, frequency, power, and space/codebook domains. 
Additionally, there are conflicting objectives to consider: minimizing outages
and maximizing capacity, which favors cell-center users. For 6G-and-beyond, operators are considering deployments in new bands while integrating/coexisting with complementary technologies, like satellite communications and Wi-Fi. 
This requires tackling large-scale combinatorial problems to maximize the network's communication and sensing capabilities without jeopardizing the performance of other incumbent services.

\subsubsection*{Classical Approaches}

Existing approaches to RAN deployment optimization rely on various techniques, each with its own limitations. The Third Generation Partnership Project (3GPP) employs global optimization methods based on stochastic system simulations. These simulations typically apply to small, homogeneous hexagonal layouts where exhaustive search techniques determine fixed parameters, such as uniform antenna downtilt angles across all cells. However, these methods do not generalize well. In real-world networks with diverse and complex configurations, site-specific radio frequency planning tools are used, relying heavily on trial-and-error methods and field measurements. These approaches are time-consuming and fail to achieve scalable and near-optimal solutions. More advanced techniques, such as RL and Bayesian optimization (BO), have been explored \cite{tekgul2023joint}. RL is able to adapt to dynamic environments but requires substantial data 
and has slow convergence rates. Also, RL lacks safe exploration and can lead to suboptimal cell configurations that degrade system performance. 
Conversely, BO offers safer exploration and faster convergence, but suffers when handling high-dimensional problems.

\subsubsection*{Graph-centric Nature}

In some cases, RAN optimization problems may be encoded as graphs.
Consider a simple cellular frequency reuse optimization problem as an illustrative example, where vertices are assigned to a cell  representing the carrier frequency allocated to it and edges connect neighboring cells. This approach can be generalized to problems involving more system parameters. For instance, vertices might represent beam configurations, edges could account for co-channel interference between cells via soft constraints, and self-edges could model the effect of each vertex on a given key performance indicator (KPI), such as sum-throughput. In this sense, the vertex value could be the angular tilt of a cell's antenna array or its allocated carrier frequency band.

\subsection*{\bf User-centric Mobility Management}

The most important mobility issues are handovers (HOs), i.e. when a user switches from one cell to another. These must be optimized preventing connectivity interruptions (late HOs) while minimizing unnecessary (early) HOs. Handovers can also be forced to rebalance traffic load and enable energy savings through opportunistic carrier shutdowns. 

\subsubsection*{Classical Approaches}

Traditionally, HOs are triggered by user-agnostic power and time thresholds. However, this one-size-fits-all approach may be unsuitable for next-generation mobile networks due to node heterogeneity. Prime use cases include air-to-ground connectivity, with aerial users experiencing much faster signal and interference fluctuations than their terrestrial counterpart. HO decisions must account for disparities in cell footprints, and, in the case of integrated terrestrial and non-terrestrial networks, capitalize on the predictable ephemeris (trajectory) of satellites. 
As a representative example, take a three-sector cell deployment with a $200$\,m intersite distance, where optimizing a single parameter per cell (e.g., the Cell Individual Offset) with five possible values over an area of $0.5$ km$^2$ results in more than $10^{40}$ possible value combinations.
Therefore, a fundamentally different approach to mobility management may be needed, with distributed, user-centric HO decisions made in real-time. These scenarios can be formulated as a Markov decision process, where deep RL emerges as a natural method.
However, a significant limitation on RL algorithms is the lack of efficient ways to learn over large, high-dimensional search spaces.

\subsubsection*{Graph-centric Nature}

We can represent the mobility management problem as a graph, with cells as vertices and connecting edges when a HO is viable. Features can be added to the vertices, with parameters such as the cell load or measurement reports. Within an RL actor-critic approach, an actor makes a HO decision for each edge given a user equipment (UE). 

\subsection*{\bf Virtual Network Function Scheduling}


Virtual Network Function (VNF)
%
%
scheduling deals with forming a 
forwarding graph for the optimal deployment and execution of interconnected VNFs, which are software implementations of network functions (such as firewalls or load balancers)
that run on virtualized infrastructure.  
The entire forwarding graph has to meet performance and resource utilization requirements, involving the current load on the physical servers, the specific resource requirements of each VNF, and the need for low latency and high throughput. 

\subsubsection*{Classical Approaches}

The scale of VNF scheduling encompasses 
the number of VNFs, the complexity of their interactions (data must flow seamlessly between VNFs while adhering to latency and bandwidth constraints), and the stringent performance and reliability requirements. Also, network conditions, resource availability, and service demands can fluctuate rapidly, requiring the VNF scheduling to be highly adaptive. All these features, make VNF scheduling highly combinatorial and complex.
In fact, an optimal allocation of VNFs is an NP-Hard problem. Proposed algorithms can be divided between exact and heuristic methods. Exact ones, based on Binary Integer Programming and Mixed Integer Linear Program, present an exponential growing complexity in problem size, making them unsuitable for data centers, edge nodes, and cloud resources with thousands of physical machines.
To improve the scalability, heuristics are introduced at the expense of optimality, usually based on iterative greedy and dynamic programming algorithms. 
Also, RL algorithms offer a promising alternative, but 
they may be inadequate to find solutions that meet service demand fluctuations and KPIs in future networks. This is due to the highly complex search over a dynamically changing, vast configuration space of large-scale distributed and heterogeneous infrastructure.


\subsubsection*{Graph-centric Nature}

The scheduling of VNFs can be represented in a graph where both the VNFs and the underlying infrastructure are modeled using nodes and edges. Within this representation, the Virtual Network Function-Forwarding Graph (VNF-FG) is a directed acyclic graph, where each node represents a VNF and each directed edge the data flow dependencies. The deployment infrastructure 
is characterized by an independent graph, where nodes correspond to physical or virtual servers and edges to network links between them.
The goal in the VNF-FG scheduling is the mapping of the VNF-FG into the infrastructure graph. Specifically, each VNF node 
needs to be assigned to a server node, 
 and each edge in the VNF-FG 
must be connected to a path in the infrastructure graph. 
There is a tradeoff in terms of resources between placing VNFs online or in batches. Once placed, the traffic needs to flow following the shortest path, which at the same time depends on the placement.

\section{Quantum Computing for Large-scale\\Mobile Network Optimization}

Once formulated as graph-centric optimization, the three classes of problems described above could be tackled with a unified strategy leveraging QC.
Indeed, in many cases the vertices of the graph may be assigned to qubits (visible and hidden variables). The energy of this system, characterized by a Hamiltonian operator, is identified with the objective function. 
Thus, 
an optimization task corresponds to finding the state of minimum energy, 
solving a combinatorial optimization problem over a high-dimensional search space, and involving Pareto optimization. 

\begin{figure}[!t]
\centering
\includegraphics[width=0.96\columnwidth]{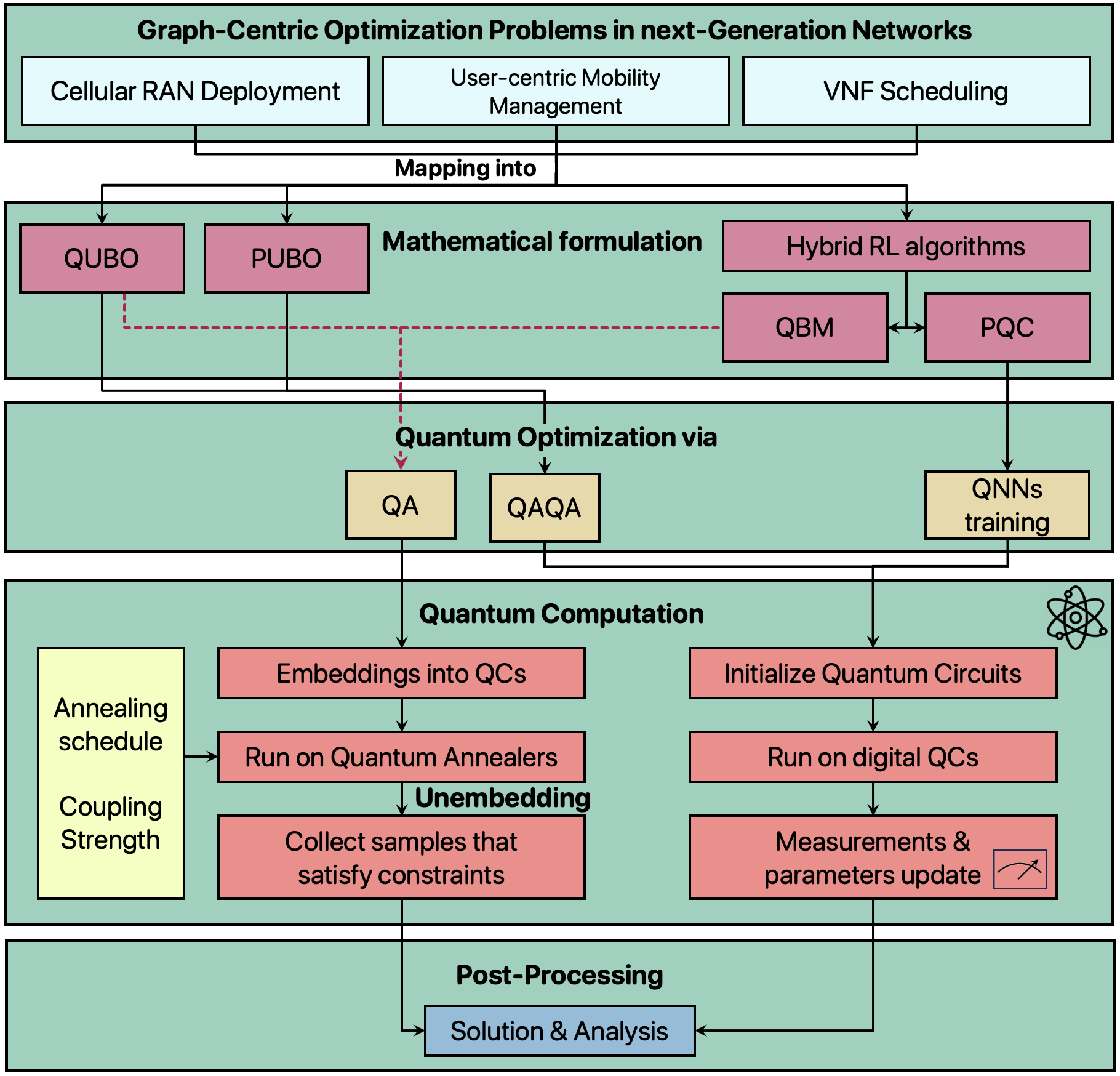}
\caption{  schematic algorithmic workflow of different QC approaches (this is not an implementation-ready pipeline). A problem is mapped to QUBO/PUBO or QML methods. Next, optimization is carried out via QA (analog QC), QAOA (digital QC), or QNNs training (digital QC).}
\label{fig:architecture-workflow}
\vspace{-0.2cm}
\end{figure}

We show in Fig. \ref{fig:architecture-workflow} a schematic workflow of different QC approaches. Initially, the problem is mathematically formulated to be amenable to QUBO, PUBO or QML methods, implemented with different QC techniques. An embedding mapping of variables into the QC architecture (physical qubits) is necessary for QA, together with hyperparameters settings such as the annealing schedule. Afterwards, samples that satisfy problem constraints are collected and analyzed. On the other hand, QAOA and QNNs run on digital QCs. Training involves sequences of measurements and parameter updates.
Depending on the problem, the solution can be evaluated offline in a quantum computer in the cloud or, as envisaged in recent works \cite{kim2021heuristic, kim2024xresqreverseannealingquantum}, computed online in centralized radio
access network (C-RAN) settings equipped with edge quantum computers. Regarding quantum programming languages and platforms, there are open-source software development libraries, such as {\it Qiskit, Pennylane} and {\it Leap}.

Table~\ref{tab:table1} summarizes the three types of problems we consider, their mathematical nature and particular challenges, the classical methods, and the potential QC solutions proposed. 
To the best of our knowledge, there is no state of the art in terms of the development
of applying QC to 6G networks.
%
\begin{table*}[t!]
\begin{center}
\caption{Classes of problems in mobile network optimization with their associated challenges and classical vs. QC approaches.}
\label{tab:table1}\vspace{-0.1cm}
\begin{tabular}{ l | l | l | l | l} 
\hline
\rowcolor{gray!30}
\textbf{Network problem} & \textbf{Mathematical problem}& \textbf{Problem challenges} & \textbf{Classical methods} & \textbf{Possible QC method}  \\ \hline
RAN deployment  & Graph coloring & Dimensionality and heterogeneity & Stochastic simulations  & QUBO/PUBO  \\\hline
Mobility management & Markov decision process & Real-time distributed decisions & User-agnostic thresholds & Hybrid actor-critic    \\\hline
VNF allocation & Job scheduling & Dynamic heterogeneous batch composition  & Greedy,  & Hybrid actor-critic  \\
 & & and allocation for computing time trade-offs & dynamic programming & \\ 
\hline
  \end{tabular}
  \end{center}
\end{table*}

\subsection*{\bf Optimization via QA and QAOA}

These techniques are most effective in NP-hard discrete  optimization, where the objective function is a polynomial. In particular, QAOA can be applied to any polynomial while QA is best suited for quadratic ones.
There is some flexibility to extend their reach when dealing with non-polynomial analytic or black-box objective functions~\cite{gabor2022approximate}.

\subsubsection*{Example Application to Cellular RAN Deployment Optimization}
%
%
As an illustrative example, consider the problem of allocating the lowest possible number of frequency channels to each cell in a network. This problem can be naturally mapped to a variation of graph coloring, i.e., one needs to assign colors (frequencies) to a vertex such that co-channel interference is minimized.
These constraints, expressed as a cost function, are subsequently mapped to a QUBO problem, efficiently approximated via QA or QAOA, which can be computed offline.
This methodology could be generalized to more complex situations, such as the optimization of the antenna tilts at each cell, to provide the best quality of service to both ground and aerial users flying along corridors. 
In these cases we can Taylor expand the cost function, map the angular tilts into a set of discrete values and carry out binary search minimizing the cost function with a quantum annealer---or a digital quantum computer, using QAOA for instance---at each iteration. 
There are still further details to specify to implement QA (or QAOA), such as the annealing schedule, coupling strength or embeddings mapping, among others.
Two related applications
involve a proof-of-concept (PoC) where TIM Telecom Italia has optimised Physical Cell Identifier planning of radio cells in 4.5G and 5G networks using QUBO (running on a D-Wave’s
2000Q quantum computer)~\cite{GSMA}, and an online optimization of a QA-based Multiple-Input Multiple-Output (MIMO) detector system which has effectively improved detection performance, achieving near-optimal throughput (over 10 bits/s/Hz)~\cite{kim2024xresqreverseannealingquantum}.

\subsection*{\bf Optimization via Quantum Machine Learning}

QC opens the possibility of using algorithms that are not efficiently simulable with just classical devices (either CPUs, GPUs or TPUs) and that may lead to an asymptotic improvement, i.e., a better scaling (for instance, from exponential to polynomial) of the total training time. 
Thus, near future developments hint toward hybrid architectures, where
GPUs handle classical machine learning pipelines (e.g., experience replay, gradient calculations) while quantum processors accelerate specific subroutines like QBMs sampling or quantum circuit optimization.
This is the kind of improvement that results, such as those reported in~\cite{Schenk_2024}, hint at and that should be explored further to better understand the benefits of applying quantum techniques in deep RL.

The hybrid actor critic RL scheme with improved learning efficiency introduced in \cite{Schenk_2024} uses a classical actor
and a quantum critic based on a clamped QBM.
The free energy of this QBM is used to approximate the reward function of the RL algorithm. To train the network, a quantum annealer is used to efficiently estimate the free energy. Alternatively, a similar approach can be adopted using parametrized quantum circuits (PQCs) on a digital quantum computer. 

An important feature of these hybrid algorithms is that 
we only need a quantum computer for training but not for deployment, making the approach more viable for near-term networks.
Also,
the main bottleneck is often computationally expensive simulators or the need to train an RL agent offline with limited data. Quantum algorithms, which enhance learning efficiency and/or require fewer samples, may significantly reduce these bottlenecks, enabling more efficient searches.

\subsubsection*{Example Application to User-centric Mobility Management}
\begin{figure*}[!t]
    \centering
    \includegraphics[width=1.0\textwidth]{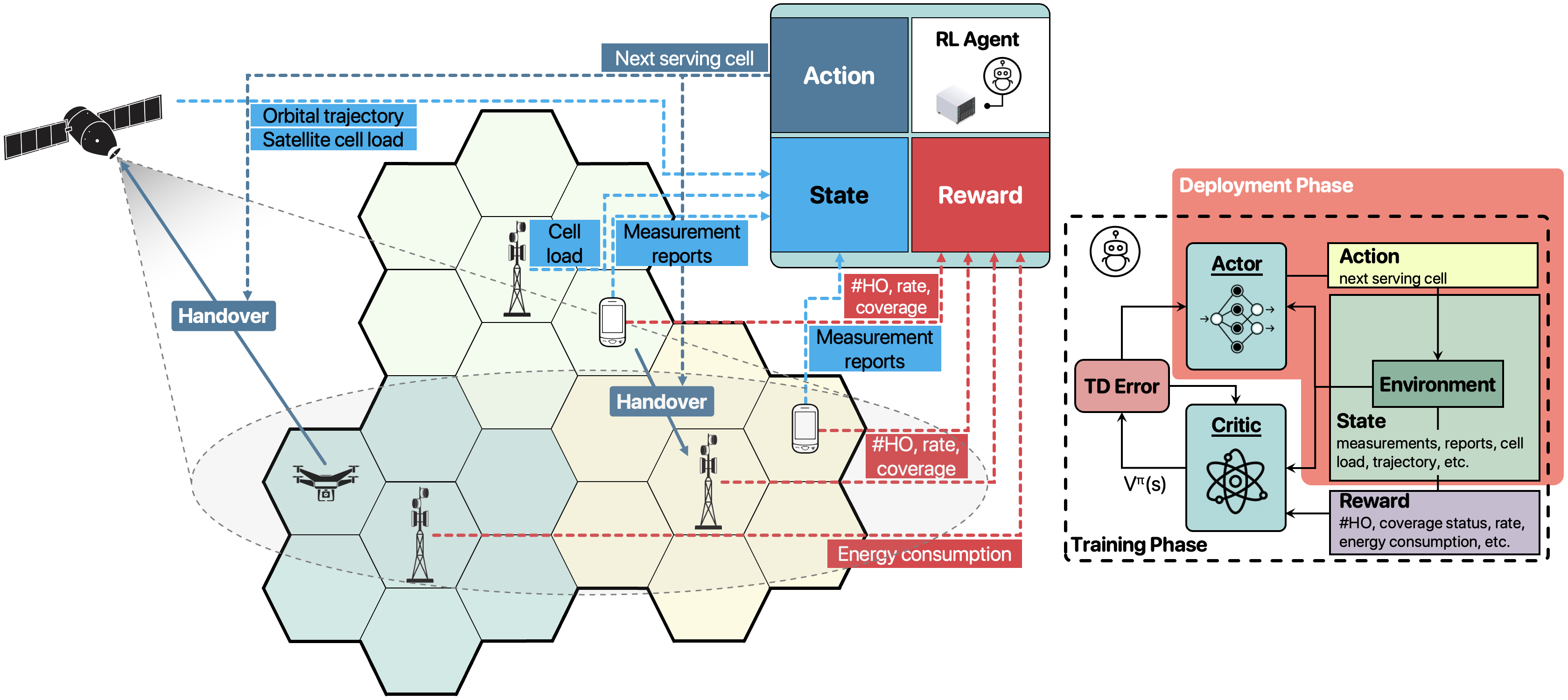}
    \label{fig:mobileHO_c}
    \vspace{-0.3cm}
    \caption{User-centric mobility management: illustration of a radio network optimizing handovers via a RL agent (left); RL agent policy training and deployment phases via QML with an actor-critic approach (right).}
    \label{fig:mobileHO}
\end{figure*}

In Fig.~\ref{fig:mobileHO}, we exemplify how a hybrid quantum-classical actor-critic RL framework can be developed for user-centric mobility management in cellular networks (a similar approach could be adapted for VNF-FG scheduling). The goal is to achieve an optimal tradeoff between minimizing the number of HOs, meeting a minimum coverage level, and maximizing the per-user data rate and the network energy savings (e.g., obtained through user offloading and carrier shutdown). 
For this problem, one could employ free energy-based RL, combining a classical policy network with a quantum-based Q-network represented by a clamped QBM. As we can see in Fig.~\ref{fig:mobileHO}, only the classical actor is required during deployment, and
the state, action and reward are defined as follows:
\begin{itemize}
\item {\it State}: features of interest, e.g. user's current serving cell, location, and direction of travel (when available), power measurement reports from available cells, cell load and, in the case of satellite cells, their ephemeris information.
\item {\it Action}: $a_s$ updates the user's serving cell.
\item {\it Reward}: weighted balance among KPIs, e.g. number of handovers, coverage status, per-user data rate, percentage of non-blocked UEs, instantaneous sum throughput, and network energy consumption.
\end{itemize}

\subsection*{\bf Scalability Considerations}

\begin{figure}[!t]
\centering
\includegraphics[width=\columnwidth]{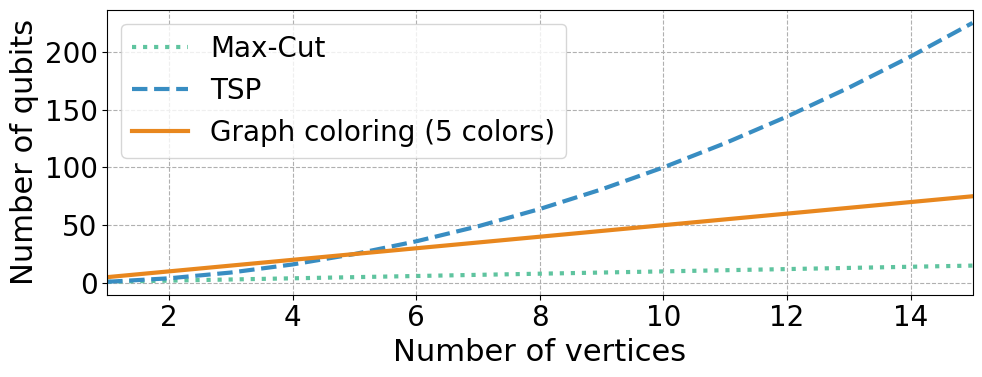}
\caption{Scalability comparison among QC solutions for Max-Cut, the traveling salesman problem, and graph coloring.  For graph coloring and Max-cut the
scaling is linear, and quadratic for the TSP.}
\label{fig:use-case}
\vspace{-0.2cm}
\end{figure}

The scalability of QC on the classes of problems presented in this paper is generally unknown. However, QC methods have been studied for classic graph problems like graph coloring, Max-Cut, and the traveling salesman problem (TSP).
The QC scalability for these three  problems, in terms of the number of graph vertices vs. qubits needed, is shown in Fig.~\ref{fig:use-case}.
This asymptotic behaviour follows directly from the standard formulations of these combinatorial optimization problems as QUBO instances that can be found, for instance, in~\cite{combarroBook}.
Max-Cut is well-suited for QUBO/Ising representations and requires a number of qubits linear with the number of vertices. (In practice, some vertices may be mapped to multiple qubits if the quantum computer topology is not dense enough.) For TSP, the required qubits grow quadratically with the number of vertices. Graph coloring is an intermediate case, with the number of qubits depending on both the number of vertices ($n$) and of colors ($k$), as $nk$. 
Sometimes, the number of qubits can be reduced if symmetries are found or with more succinct QUBO representations.

Overall, qubit scaling varies from linear to at most quadratic. 
For QUBO approaches in RAN optimization, typically based on adaptations of graph coloring (e.g., adding soft couplings to model interference, relaxing constraints), we generally expect sub-quadratic scaling. For the use cases better suited for QML, scaling partially decouples from the number of vertices. For instance, with clamped QBMs, we would need $m$ qubits per vertex (where $m$ is the sum of the state space and action dimensionality) for the self-couplings between visible and hidden variables~\cite{Schenk_2024}, and additional qubits for each layer and hidden variable in the model.

\section{Challenges and Research Directions} \label{sec:challenges}

We now discuss the key challenges 
for QC to move from a promising technology to a practical tool driving the large-scale optimization of future mobile networks.

\subsection*{\bf Algorithmic Aspects}\label{algorithmic-challenges}


\subsubsection*{Novel Algorithms}

A significant challenge is developing novel QC algorithms to deal with larger and more realistic problems. Promising directions include custom-made solutions to effectively exploit NISQ hardware,
as well as 
new architectures for QML.

\subsubsection*{Mathematical Formulations}

A  challenge is creating efficient and realistic mathematical problem formulations, suitable for QA/QAOA and QML. Current implementations are typically simplified proof-of-concepts. 

\subsubsection*{Mapping to Quantum Embeddings}

To run an algorithm, classical variables must be mapped to the quantum state space, i.e., physical qubits. 
Custom-designed QC embeddings are crucial for performance and scalability, leveraging the limited power of current NISQ devices.


\subsection*{\bf Architecture Aspects}

\subsubsection*{Architectural Integration}
The integration of quantum and hybrid optimization workflows in the structure of beyond-6G networks involves several interrelated variables, including the requirements of each use case in terms of QC power and time-to-solution, the location of the quantum hardware, and its distance from the sources of data. Most of the examples discussed in the paper can be computed offline, therefore using a quantum computer deployed in the cloud, because either the nature of the problem (e.g. frequency allocation) or of the solution (e.g. hybrid RL with QC used during training only) allows it. However, latency-sensitive optimization problems solved online with QC solutions, such as  \cite{kim2024xresqreverseannealingquantum}, will require the deployment of QC closer to the data, possibly in a C-RAN fashion. Fortunately, rack-mountable quantum computers have appeared recently. Regardless, the architectural challenge remains in the development of means of abstraction and integration that consider both sides to achieve both a near-optimal placement of QC resources and a schedule of the optimization requests (from handovers, VNF scheduling, and other functions) that satisfy the requirements of the network. The orchestration of quantum-classical computing systems is a recent yet active area of research \cite{faro2023middleware}, but solutions are generic and should be streamlined to the particularities of 6G-and-beyond networks (e.g. distributed data, stringent time-to-solution requirements).

\subsubsection*{Algorithm Compilation}
To cope with the fragile nature of qubits, quantum compilers need to be surgical to minimize the circuit depth and remove wasteful operations, which is challenging due to the many variables involved.
To this end, compilers need to be algorithm-specific, adapting their optimizations to the input data (graph). They also should be hardware-aware, adapting their operations to calibration data.
\subsubsection*{Practicability of methodology}
To assess the practicability of our proposed methods, performance evaluation results with testbeds at the simulation or experimental level are needed.

\subsection*{\bf Quantum Computing Aspects}


\subsubsection*{NISQ Devices and Error Correction} 

In quantum gate computers, such as those from IBM and Google, estimates range from 100 to 1000 physical qubits per error-corrected logical qubit. For quantum annealers, noise is less problematic but 
sparse connectivity requires representing logical qubits with sets of physical qubits (embeddings). 

\subsubsection*{Scaling-up Quantum Computers}

To scale QC, several groups have proposed to interconnect multiple quantum processors \cite{rodrigo2021double}. This has multiple implications. For instance, the mapping of the quantum algorithms needs to be adapted to minimize qubit exchanges across processors, which is challenging. It also opens an opportunity to match a distributed architecture with the distributed nature of certain 6G optimization problems.

\section{Conclusion} 
\label{sec:conclusion}

This paper examined the potential of quantum computing to advance large-scale optimization in future mobile networks. While achieving quantum advantage in practical combinatorial tasks remains unresolved, we offered a structured vision for integrating quantum computing into network optimization, identifying critical challenges and research directions necessary for its practical deployment.

\bibliographystyle{IEEEtran}
\bibliography{IEEEabrv,bib}

\section*{} 
\label{sec:bios}

\vspace{-1cm}

\begin{IEEEbiographynophoto}{Sebastian Macaluso} is a Principal Research Scientist at Telefónica. His research interests include innovative algorithms on quantum computing, deep learning and statistical inference.
\end{IEEEbiographynophoto}
\vskip -2\baselineskip plus -1fil
\begin{IEEEbiographynophoto}{Giovanni Geraci} is a Principal Research Scientist at Telefónica and an Associate Professor at Universitat Pompeu Fabra (UPF) in Barcelona. He was with Nokia Bell Labs and served as an IEEE ComSoc Distinguished Lecturer.
\end{IEEEbiographynophoto}
\vskip -2\baselineskip plus -1fil
\begin{IEEEbiographynophoto}{Elías F. Combarro} 
is a full professor at the University of Oviedo. His research interests include quantum machine learning and the application of quantum algorithms to algebraic problems.  \end{IEEEbiographynophoto}
\vskip -2\baselineskip plus -1fil
\begin{IEEEbiographynophoto}{Sergi Abadal} 
is a Distinguished Researcher at Universitat Polit\`{e}cnica de Catalunya (UPC). His research interests include quantum computing, wireless communications, and computer architecture.
\end{IEEEbiographynophoto}
\vskip -2\baselineskip plus -1fil
\begin{IEEEbiographynophoto}{Ioannis Arapakis} 
 is Director of Telef\'{o}nica Scientific Research and Adjunct Professor at the Barcelona School of Economics. His research interests include deep learning, information retrieval and cognitive science.
 \end{IEEEbiographynophoto}
\vskip -2\baselineskip plus -1fil
\begin{IEEEbiographynophoto}{Sofia Vallecorsa} is a researcher in the fields of Artificial Intelligence and Quantum Computing with applications in High Energy Physics at CERN. She is the coordinator of the CERN Quantum Technology Initiative. \end{IEEEbiographynophoto}
\vskip -2\baselineskip plus -1fil
\begin{IEEEbiographynophoto}{Eduard Alarc\'{o}n} is full professor at the Universitat Polit\`{e}cnica de Catalunya (UPC). His research interests include communications at the nano-scale, wireless energy transfer and computer architecture.
\end{IEEEbiographynophoto}

\end{document}